\documentclass{aa}  

\usepackage[T1]{fontenc}
\usepackage{hyperref}
\usepackage{changes}
\usepackage[section]{placeins}
\usepackage[titletoc]{appendix}
\usepackage{xcolor}
\usepackage{dsfont}
\usepackage{bm}
\usepackage{mathtools}
\usepackage{enumerate}
\usepackage{braket}
\usepackage{scalerel} 
\usepackage{subeqnarray}
\usepackage{dutchcal} 
\usepackage{comment}

\usepackage{graphicx}	% Including figure files
\usepackage{float} 
\usepackage{amsmath}	% Advanced maths commands
\usepackage{amssymb}
\usepackage{gensymb} 
\usepackage{lineno}

\DeclareMathAlphabet{\mathpzc}{OT1}{pzc}{m}{it}

\newcommand{\sayy}[1]{`#1'}
\DeclarePairedDelimiter\abs{\lvert}{\rvert}%
\def\be{\begin{equation}}
\def\ee{\end{equation}}
\def\bea{\begin{eqnarray}}
\def\eea{\end{eqnarray}}

\newcommand{\obs}{{\rm o}}%observation
\newcommand{\src}{{\rm s}}%source
\newcommand{\dd}{{\rm d}}

\definecolor{MyB}{rgb}{0.1,0.1,1.0}

\begin{document}

\title{Position drift with Gaia}

\author{Andreas~Tsigkas~Kouvelis\inst{1}
        \and Asta~Heinesen\inst{1}\fnmsep\thanks{Corresponding author: A.~Heinesen\newline
e-mail: asta.heinesen@nbi.ku.dk} \and Shashank~Shalgar\inst{1} \and Miko\l{}aj Korzy\'nski\inst{2}}

\institute{Niels Bohr Institute, Blegdamsvej 17, DK-2100 Copenhagen, Denmark 
            \and  Center for Theoretical Physics, Polish Academy of Sciences, Al. Lotnik\'o{}w 32/46, 02-668 Warsaw, Poland\\ }

\date{}

\abstract 
{The proper motion (also known as position drift) field of extragalactic sources at cosmological distances across our sky can be used to measure the acceleration of the Solar System through the aberration effect. 
If measured very precisely, the signal would also hold cosmological information, for instance about bulk flows of distant sources or the presence of tensor modes.}
{In the $\Lambda$ cold dark matter ($\Lambda$CDM) model, the acceleration of the Solar System is by far the dominant contributor to the position drift signal for sources at cosmological distances, and the measurement is therefore expected to yield a constant spheroidal dipole across redshifts as long as convergence to the cosmic restframe has been reached. The aim of this paper is to test this hypothesis.  } 
{We analyze data from the cosmic reference frame (CRF) dataset of Gaia data release 3 (DR3) focusing on constraining the dipole and quadrupole in the position drift signal, with an emphasis on redshift dependence of the signal as a consistency test of the $\Lambda$CDM model.} 
{The spheroidal dipole that we find is in mild tension, at the level of $2-3\sigma$, with the constant-in-redshift signature expected from the local acceleration of the Solar System. We also find significant quadrupole components, that however do not have any significant evolution with redshift.  The most straightforward interpretation of these findings is (unknown) systematic errors related to the Gaia instrumentation, but a cosmological origin is a possibility.} 
{Our analysis remains inconclusive on the cause of the redshift dependence of the dipole and warrants further investigations with upcoming data releases. We discuss possible implications of our results and highlight the importance of proper motion measurements for rest frame determinations in cosmology. 
In our discussion, we highlight interesting avenues for doing cosmology with Gaia data.}


\maketitle

\nolinenumbers

\section{Introduction} 
Real-time cosmology is an exciting new field that is concerned with following sources over time using telescopes in order to directly extract information about the expansion of the Universe as well as local kinematics. 
The most examined observables in real-time cosmology are the redshift drift and the position drift, which are the time evolutions of redshifts and sky-positions, respectively, of sources. 
While the measurement of redshift drift is still futuristic, although within reach within the next few decades, the position drift signal has already been measured. The first measurements were carried out using Very Long Baseline Interferometry (VLBI) techniques \cite{2012A&A...544A.135X,2011A&A...529A..91T, 2012ivs..conf..352X}. More recently, the Gaia spacecraft measured the position drift by tracking the positions of extragalactic sources in the optical band \cite{2022A&A...667A.148G, 2021A&A...649A...9G}.

The proper motions of extragalactic objects across our sky are estimated by Gaia with errors $\gtrsim \text{mas/yr}$ for the individual objects, with a mean that is determined with an accuracy of $\sim \mu\text{as/yr}$ with the $\sim 1$ million quasi-stellar objects (QSOs) in the cosmic reference frame (CRF) dataset {contained in the} Gaia Data Release 3 (DR3) \cite{2022A&A...667A.148G}. 
This precision is enough to determine the acceleration of the Solar System \cite{2021A&A...649A...9G}. 

{The position drift of sources at cosmological distances across our sky contains rich additional information that could in principle be extracted by an accurate enough estimate of the proper motions. 
This includes information on the Solar System's peculiar velocity \citep{Paine:2019vep}, the gravitational wave background \cite{gwenergydensity}, intrinsic cosmological shear (such as that present in Bianchi space-times) \citep{Marcori:2018cwn}, and vorticity \cite{2012ApJ...755...58N,Amendola:2013bga}. 
Position drift therefore also holds the promise to test the $\Lambda$CDM framework in various ways, hereunder to test the extent to which the cosmological principle applies \cite{2022ApJ...927L...4M}. 

In the $\Lambda$CDM model, the local peculiar acceleration of the Solar System relative to the Poisson frame is by far the dominant source of the signal. 
Since the inference of the local acceleration of the Solar System should be consistent across different datasets, a simple test of the $\Lambda$CDM model prediction is to test the consistency of the position drift signal across datasets of different cosmological redshifts. 
Such consistency tests have been relatively little explored, but were  considered for VLBI sources by \citet{2011A&A...529A..91T} as well as for Gaia sources in recent analyses by \citet{2022ApJ...927L...4M,Makarov:2025vtb}. 
In modifications of the $\Lambda$CDM model, the signal could acquire additional terms on top of the acceleration of the Solar System, and a significant redshift dependence could thus ultimately be a sign of new physics. 

In this paper, we perform such a test using the CRF dataset of Gaia DR3, by dividing it into smaller datasets based on the redshifts of the sources.  
We focus on the redshift dependence of the spheroidal component of the dipole in the position drift distribution of the QSOs (which is the component that is conventionally interpreted as the peculiar acceleration of the Solar System), but we also consider the rotation of the sources across the sky as well as the quadrupole of the position drift signal. 

In section \ref{sec:setup}, we describe the theoretical prediction of the position drift signal within FLRW models with peculiar motions of sources on top.

}

\vspace{5pt} 
\noindent
\underbar{Notation and conventions:}
Units are used in which $c=1$.
The signature of the space-time metric $g_{\mu \nu}$ is $(- + + +)$ and the connection $\nabla_\mu$ is the Levi-Civita connection.
Greek letters $\mu, \nu, \ldots$ label space-time
indices in a general basis, running from 0 to 3. 
{We also occasionally use bold notation $\bm V$ for vectors in basis free notation.}

\section{Theory of position drift in FLRW spacetimes}
\label{sec:setup}
%%%%%%%%%%%%%%%%%%%%%%%%%%%%%%%%%%%%%%%%%%%%%%%%%%%%%%%%%%%%%%%%%%%  
Consider the Friedmann-Lema\^{\i}tre-Robertson-Walker (FLRW) line element 
\begin{equation}
\dd s^2 = -\dd t^2 + a^2(t) ( \dd \chi^2 + R_0^2\,S_k(R_0^{-1}\,\chi)^2 \dd \Omega^2  )  \, ,  %\dd s^2 = -\dd t^2 + a^2(t)\delta_{ij}\dd x^i \dd x^j
\end{equation} 
in units where $c=1$, where $t$ is the proper time of the comoving observers with 4-velocity $u^\mu\equiv \delta_0^\mu$, the radial coordinate $\chi$ has the dimension of length, $R_0$ is the curvature radius, also of dimension of length, and $a(t)$ is the dimensionless scale factor. As usual, we fix $a(t)$ to be 1 at the present moment $t=t_0$. The dimensionless parameter $k$ governs the sign of the spatial curvature and it may thus take one of three possible values: $0, \pm 1$. 
The metric on the unit sphere is 
\begin{equation} \label{angularel}
\dd \Omega^2 = \dd \delta^2 + \cos^2 \! \delta \,\dd \alpha^2
\end{equation} 
where\footnote{The declination angle is defined as $\delta \equiv \pi/2 - \delta_{\text{inc}}$ where $\delta_{\text{inc}}$ is the usual inclination angle of an spherical coordinate system.} $\delta$ represents the declination angle and $\alpha$ represents the right ascension angle in the equatorial coordinate system. 
We use here the standard notation $S_k$ denoting the function
\begin{equation}
\label{Sk}
S_k(\chi) =\begin{cases}
    \sin \chi & \text{if $k = 1$} \\
    \chi & \text{if $k=0$} \\
    \sinh \chi & \text{if $k = -1$}
\end{cases}.
\end{equation} 
We also introduce $C_k(\chi)$ as the derivative of \eqref{Sk}, yielding 
\begin{equation}
\label{Ck}
C_k(\chi) = \begin{cases}
    \cos \chi & \text{if $k = 1$} \\
    1 & \text{if $k=0$} \\
    \cosh \chi & \text{if $k = -1$}
\end{cases}.
\end{equation} 
We let the conformal time be defined by $\dd\eta=a^{-1}\dd t$ and the Hubble function by $H (t)\equiv \dd\ln a/\dd t$, with the Hubble constant $H_0 \equiv H(t_0)$ being the Hubble parameter function as evaluated at the present epoch $t=t_0$. 
Recall also that the curvature radius $R_0$ 
sets the scale of the spatial curvature and is related to the dimensionless curvature parameter $\Omega_{k,0}$, $k$ and the Hubble constant $H_0$ 
via $\Omega_{k,0}  = - k/ (H_0^2\,R_0^2)$.

We consider observers and sources in motion relative to the comoving observers, such that they may have velocities relative to $u^\mu$ that are not restricted in any way other than that they have to be much smaller than the speed of light. We may then write their 4-velocities, up to the linear order in relative velocity, as 
\begin{equation}
\label{eq:utilde}
\tilde{u}^{\mu}=u^{\mu}+v^{\mu},
\end{equation} 
where $v^{\mu}$ is the relative velocity satisfying $v^\mu u_\mu=0$, 
and where tilted variables from now on refer to the observers/sources in motion relative to the comoving frame. 

We consider a spatial direction vector $n^\mu$ to an object in the frame of $u^\mu$, such that $n^\mu u_\mu=0$. 
In the frame of the moving (tilted) observer, we have 
\begin{equation}\label{ntilde}
\tilde{n}^{\mu} = n^{\mu}+\bot^{\mu}_{\nu}\,v^{\nu} + v^\sigma\,n_\sigma\,u^\mu,
\end{equation} 

where the projection tensor $\bot^{\mu}_{\nu}$ is defined as 
\begin{equation}\label{projection}
\bot^{\mu}_{\nu} \equiv \delta^{\mu}_{\nu} + \tilde{u}^{\mu} \tilde{u}_\nu  - \tilde{n}^{\mu} \tilde{n}_\nu  \,  
\end{equation} 
and the last term in \eqref{ntilde} ensures that $\tilde n_\mu\,\tilde u^\mu = 0$, i.e. $\tilde n^\mu$ is spatial with respect to the tilted source or observer. 
We are interested in the drift of the direction vector \eqref{ntilde} for source on the sky as a function of observer proper time, which we shall also refer to as proper motion of the source. It is mathematically defined in \emph{any} given space-time geometry as \cite{Heinesen:2024npe}  
\begin{equation}\label{mudef}
\mu^{\alpha} \equiv \bot^{{\alpha}}_{\nu} \,\tilde{u}^\sigma \nabla_\sigma \tilde{n}^\nu  \, , 
\end{equation} 
where it is implicit that the derivative is performed for every individual source, so that $\tilde{n}^\nu$ as evaluated along the observer worldline traces the motion across the sky of that source. 
For the FLRW model with relative velocities considered in this section, the proper motion \eqref{mudef} can be written in terms of the relative velocity field and the FLRW geometrical quantities as 
\begin{equation} \label{ntildedrift}
%\frac{\delta  \tilde{n}^{\mu} }{ \delta \tilde{t}_\obs } 
\mu^{{\alpha}} = \bot^{{\alpha}}_{\nu}\,\dot v_\obs^\nu + \frac{\bot^{{\alpha}}_{\nu}\,v^{\nu}_\src}{R_0\,S_{k}(R_0^{-1}\,\chi)}-\frac{C_k(R_0^{-1}\,\chi)}{R_0\,S_k(R_0^{-1}\,\chi)}\,\bot^{{\alpha}}_{\nu}\,v^{\nu}_\obs ,
\end{equation} 
where we have defined 
\begin{equation} 
    \dot v_{\obs}^{\mu} \equiv \tilde{u}^\nu \nabla_\nu v_{\obs}^{\mu}  = \dot{\tilde{u}}_{\obs}^{\mu} - H_0\,v_{\obs}^\mu %+  O(v_\obs^2), %= \nabla_{\tilde u} v_{\obs}^\mu +
\end{equation}
where $\dot{} \equiv \tilde{u}^\mu \nabla_\mu$ 
denotes the covariant derivative along the observer's worldline. 
It follows that $\dot{\tilde{v}}_{\obs}^{\mu}$ is the total 4-acceleration of the observer, consisting of the non-gravitational 4-acceleration $ \dot{\tilde{u}}_{\obs}^{\mu} \equiv \tilde{u}^\nu \nabla_\nu \tilde{u}^\mu$ and the Hubble drag term. 
{In cosmological perturbation theory $\dot{\tilde{v}}_{\obs}^{\mu}$ has interpretation as the peculiar acceleration of the observer relative to the Poisson frame.}  

The expression in \eqref{ntildedrift} reduces to that found in equation~(3.14) of \citep{Marcori:2018cwn} in the spatially-flat FLRW case. We note that the presence of non-zero curvature generally breaks the symmetry between the $v^{\nu}_\src$ term and the $-v^{\nu}_\obs$ term in \eqref{ntildedrift} due to the curvature factor $C_k(R_0^{-1}\,\chi)$ multiplying the latter term. 

{In a full treatment within cosmological perturbation theory, the expression \eqref{ntildedrift} is recovered as the dominant contribution (on top of which there are generally effects from the integrated Sachs Wolf effect, gravitational redshift, and tensor modes). At cosmological distances, the peculiar acceleration term entering as the first term in \eqref{ntildedrift} is furthermore expected to be several orders of magnitude larger than the two remaining peculiar-velocity terms in the expression for sources for realistic values of the Solar System acceleration, due to the scaling of the peculiar-velocity terms with inverse distance. Thus, when position drift data is interpreted within the $\Lambda$CDM model, the prediction of the signal in practice reduces to the peculiar-acceleration term.  }

\section{Method}

\subsection{Data}
Following \cite{2021A&A...649A...9G}, we use a set of 1,215,942 five-parameter QSO-like sources from the cosmic reference frame (CRF) dataset of the Gaia DR3 \cite{2022A&A...667A.148G}{, which we shall also refer to as CRF3.} 
Each source has a five-parameter astrometric solution consisting of positions $(\alpha, \delta)$, parallax $(\varpi)$ and proper motions $(\mu_{\alpha*}, \mu_{\delta})$, 
{which are defined as the time-changes of the positions\footnote{The  cosine in the definition comes from the form of the metric on the unit sphere~\eqref{angularel}.}  $\mu_{\alpha*} \equiv \mu_{\alpha} \cos\delta = \dot{\alpha} \cos\delta$ and $\mu_{\delta} \equiv \dot{\delta}$}.  
These sources are identified as extragalactic through cross-matching with external QSO and AGN catalogs and are selected for their negligible intrinsic proper motions compared to other types of sources in the Gaia catalog.

\begin{figure}
\includegraphics[width=\columnwidth]{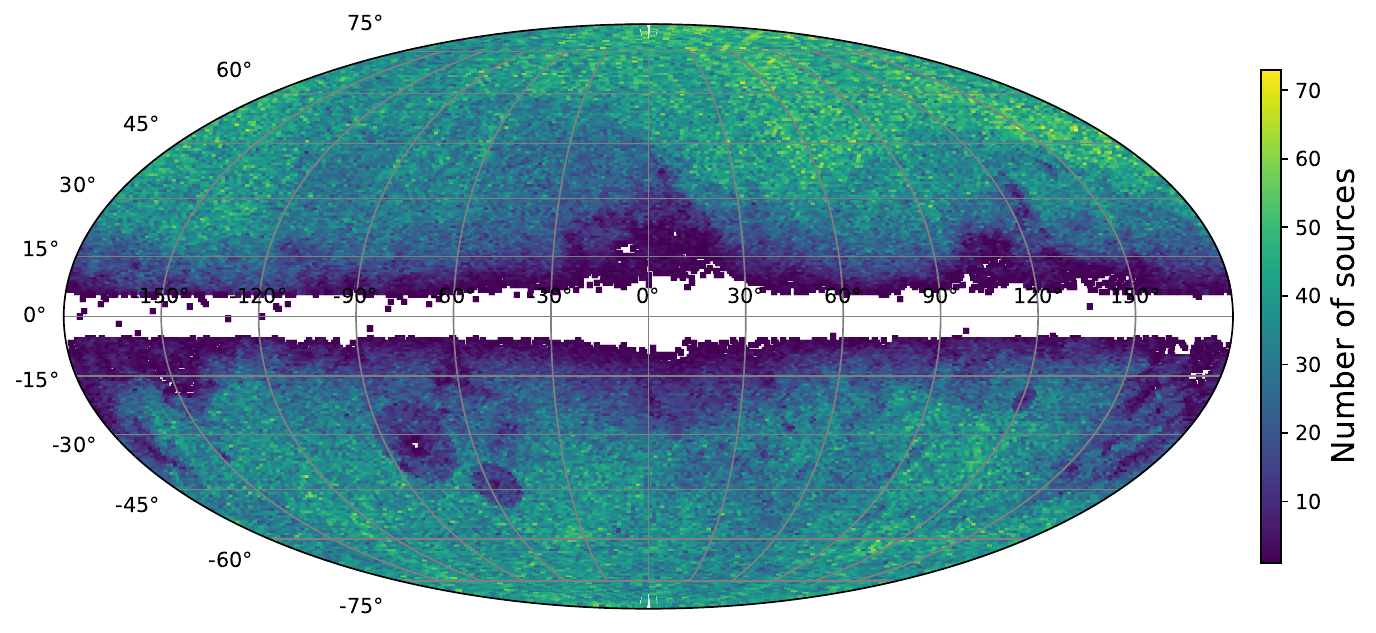}
\caption{Sky distribution of the 1,215,942 five-parameter QSO-like sources from the CRF dataset of Gaia DR3.}
\label{fig:sky_distribution}
\end{figure}

{The sky distribution of the sources is shown in Fig.~\ref{fig:sky_distribution}.} 

\begin{figure}
\includegraphics[width=\columnwidth]{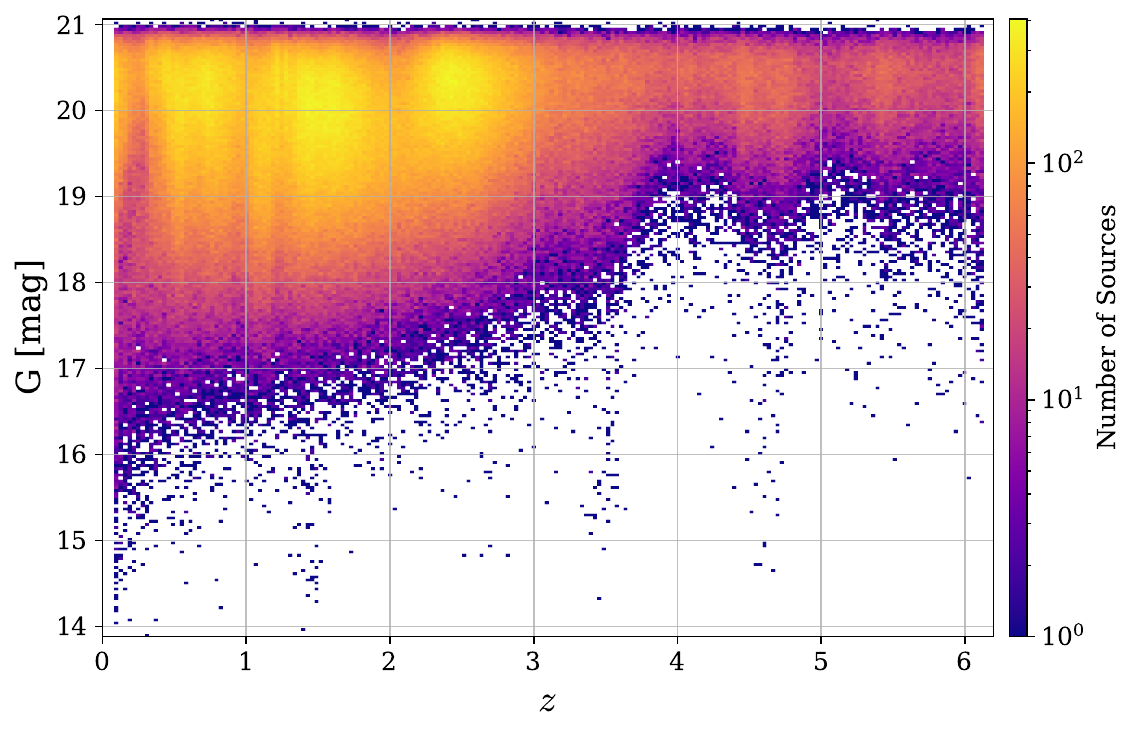}
\caption{Distribution of the 1,170,933 Quasi Stellar Object Classifier identified sources in terms of their redshifts and G-band magnitudes.}
\label{fig:z_distribution}
\end{figure}

For the {redshift determinations of the same sources}, we use that provided by the Quasi Stellar Object Classifier\footnote{See also the \href{https://gea.esac.esa.int/archive/documentation/GDR3/Data_analysis/chap_cu3qso/sec_cu3qso_char/ssec_cu3qso_char_redshift.html}{Gaia data release 3 documentation on redshifts}.} \cite{Delchambre:2022ugo}{, with which we have redshift information for 1,170,933 of the sources available} ranging from  $z = 0.0826$ to $z=6.12295$. 
The distribution of sources in redshift and in G-band magnitude is shown in Fig.~\ref{fig:z_distribution}. 

\subsection{Vector Spherical Harmonics} 
We decompose the proper motion field into a series of vector spherical harmonics (VSH) defined on the celestial sphere, following the formalism of \citet{2012A&A...547A..59M}.  The proper motion at each position is treated as a tangent vector field on the sphere, written as $\bm{\mu}(\alpha, \delta) = \mu_{\alpha*} \bm{\mathcal{e}}_{\alpha} + \mu_{\delta} \bm{\mathcal{e}}_{\delta}$, where $\bm{\mathcal{e}}_{\alpha} $ and $\bm{\mathcal{e}}_{\delta}$ are the local unit vectors in the directions of right ascension and declination, respectively. 
Any complex vector field defined on the sphere can be uniquely decomposed into a sum of VSHs
\begin{equation} 
\label{eq:VSH}
    \bm{\mu}(\alpha, \delta) = \sum_{l=1}^{\infty} \sum^{l}_{m = -l} \left( t_{lm} \bm{T}_{lm} + s_{lm} \bm{S}_{lm}  \right),
\end{equation}
where $\bm{T}_{lm}$ and $\bm{S}_{lm}$ are the toroidal (or magnetic) and spheroidal (also referred to as poloidal or electric) harmonics of degree $l$ and order $m$, and the corresponding coefficients $\bm{t_{lm}}$ and $\bm{s_{lm}}$ are obtained by fitting to the data. 
The spheroidal terms in \eqref{eq:VSH} constitute the curl-free part of $\bm{\mu}$ in its Helmholtz decomposition, and the toroidal terms constitute the divergence-free part \cite{1985EJPh....6..287B}.   

Since the observed field is real-valued, we can use the symmetry properties of the expansion to rewrite it as 
\begin{align}
\label{eq:mudecompositon}
    & \bm{\mu}(\alpha, \delta) = \sum_{l=1}^{\infty} \Big( t_{l0} \bm{T}_{l0} + s_{l0} \bm{S}_{l0}  \notag  \\ 
     &   \hspace{-0.2cm} +  2\sum^{l}_{m = {1}}\left(\ t_{lm}^{\Re} \bm{T}_{lm}^{\Re} - t_{lm}^{\Im}\bm{T}_{lm}^{\Im} + s_{lm}^{\Re} \bm{S}_{lm}^{\Re} - s_{lm}^{\Im} \bm{S}_{lm}^{\Im}\right)\Big),
\end{align}
where $\Re$ and $\Im$ indicate the real and imaginary parts of the coefficients and basis functions. 

The goal of this work is to estimate the peculiar acceleration of the Solar System. In the VSH framework, the $l=1$ terms contain two global features: a rotation, given by the toroidal harmonics $\bm{T_{1m}}$ and a glide, represented by the spheroidal harmonics $\bm{S_{1m}}$. The glide has the same mathematical form as the signal caused by acceleration, so by fitting the $\bm{s_{1m}}$ coefficients, we can determine the direction and amplitude of the acceleration vector.

In a constraint analysis, we must in practice truncate the outer sum in \eqref{eq:mudecompositon} by a finite $l_{\max}$ value. 
In our analysis, we include terms up to degree $l_{\max}=4$, corresponding to the dipole, quadrupole, octupole, and hexadecapole components. 
The choice to include {multipoles} at least up to $l=3$ is motivated by {the} results in  \cite{2021A&A...649A...9G} (see {their} Fig.~7), which show that the acceleration estimates become stable at this level. 
Higher-order {harmonics} have a negligible effect on the dipole signal. Nevertheless, we {conservatively} include the $l=4$ harmonics to ensure {that our estimates remain robust for} the quadrupole coefficients. 
We also include the correlations between the proper motion components for each source, and as already shown by Gaia’s analysis, these correlations change the acceleration estimates only by about 10\% of their uncertainties and therefore have a minimal impact on our results.

\subsection{Fitting the data}

We estimate the VSH coefficients by performing a Bayesian fit using the Python library PyMC and sampling with the NUTS sampler, a gradient-based MCMC algorithm. We run eight chains with 2000 tuning steps and 2000 draws each and verify the convergence using the Gelman-Rubin statistic {\cite{10.1214/ss/1177011136} $\hat{R}$ (we apply a conservative threshold of $\hat{R} < 1.001 $)} and visual inspection of the sampling traces.

We fit all VSH terms up to degree $l=4$, including both spheroidal and toroidal components. The number of parameters increases with the maximum degree $l_{\max}$ of the expansion. There are 6 coefficients for $l=1$, 16 for $l=2$, 30 for $l=3$ and 48 for $l=4$.

For the first degree, the field is described in terms of the glide vector $\bm{{a}} = ({a}_{x}, {a}_{y}, {a}_{z})$ and the rotation vector $ \bm{R} = (R_{x}, R_{y}, R_{z}) $, both defined in equatorial coordinates $ (\alpha, \delta) $. The corresponding proper motion field is given by
\begin{align} 
\label{eq:dipoletransformation}
\hspace{-0cm}    \mu_{\alpha*} {\text{(dipole)}} = &-{a}_{x} \sin\alpha + {a}_{y} \cos\alpha \notag\\
    & +R_{x} \sin\delta\cos\alpha + R_{y} \sin\delta\sin\alpha - R_{z} \cos\delta \notag \\ 
    \mu_{\delta}  {\text{(dipole)}}  = &-{a}_{x} \sin\delta\cos\alpha - {a}_{y} \sin\delta\sin\alpha + {a}_{z} \cos\delta\notag\\
    &-R_{x} \sin\alpha + R_{y} \cos\alpha \, .  \qquad 
\end{align}
{T}he quadrupole contribution to the proper motion is expressed as
\begin{align} 
\label{eq:predictedp}
    \hspace{-2cm} \mu_{\alpha*} {\text{(quadrupole)}} = \ &t_{20} \sin 2\delta\notag\\
    -&\left(t_{21}^{\Re}\cos\alpha - t_{21}^{\Im}\sin\alpha\right)\cos 2\delta\notag\\
    +&\left(s_{21}^{\Re}\sin\alpha + s_{21}^{\Im}\cos\alpha\right)\sin\delta\notag\\
    -&\left(t_{22}^{\Re}\cos 2\alpha - t_{22}^{\Im}\sin 2\alpha\right)\sin 2\delta\notag\\
    -&2\left(s_{22}^{\Re}\sin 2\alpha + s_{22}^{\Im}\cos 2\alpha\right)\cos\delta \notag\\
    \mu_{\delta} {\text{(quadrupole)}} = \ &s_{20} \sin 2\delta\notag\\
    -&\left(t_{21}^{\Re}\sin\alpha + t_{21}^{\Im}\cos\alpha\right)\sin\delta\notag\\
    -&\left(s_{21}^{\Re}\cos\alpha - s_{21}^{\Im}\sin\alpha\right)\cos 2\delta\notag\\
    +&2\left(t_{22}^{\Re}\sin 2\alpha + t_{22}^{\Im}\cos 2\alpha\right)\cos \delta\notag\\
    -&\left(s_{22}^{\Re}\cos 2\alpha - s_{22}^{\Im}\sin 2\alpha\right)\sin 2\delta.
\end{align}
Using these, together with the $l=3$ and $l=4$ terms in \eqref{eq:mudecompositon}, cf. the full list of harmonics used in the fit up to degree $l_{\max}=4$ in Appendix \ref{Appendix A}, we arrive at a prediction for the $\bm{\mu}$ field at large angular scales. 
Following \cite{2021A&A...649A...9G}, we assume a Gaussian likelihood\footnote{{See Fig.~6 of \cite{2021A&A...649A...9G}, which shows excellent agreement of the Gaussian assumption with the data.}} {for the measured proper motions $\bm{\mu}_{i}^{obs}$ where the index $i$ labels the source. The full expression for the log-likelihood is thus} 
\begin{align}
    \log \mathcal{L} = &- n \log(2\pi) - \frac{1}{2}\sum_{i=1}^{n}\log(\det(\bm{\mathcal{C}}_{i})) \notag\\&- \frac{1}{2}\sum_{i=1}^{n}\left(\bm{\mu}_{i}^{obs} - \bm{\mu}_{i}^{model}\right)^{T}\bm{\mathcal{C}}^{-1}_{ii}\left(\bm{\mu}_{i}^{obs} - \bm{\mu}_{i}^{model}\right),
\end{align}
{where $\bm{\mu}_{i}^{model}$ is the theoretical prediction for the i'th source given by \eqref{eq:mudecompositon} as truncated at a given $l_{\max}$,} {$\bm{\mathcal{C}}_{ii}$ is the $2 \times 2$ dimensional covariance matrix for the i'th source,} and $n$ is the number of sources in the sample. 
{Note that the covariance matrix does not contain covariances between different sources.} 

The parameters are then estimated for each choice of $l_{\max}$, and the results are presented in the next section.

The expected magnitude of the Solar System's acceleration is on the order of a few microarcseconds per year, based on theoretical predictions from Galactic dynamics. We therefore assign independent Gaussian priors to all coefficients with mean zero and standard deviation $\sigma = 10^{-2} \,\text{mas/yr}$. This prior choice is broad enough to not bias the fit while remaining consistent with the expected value. We tested significantly broader priors (e.g. $\sigma = 10^{2} \,\text{mas/yr}$) and found that the resulting posteriors remained stable, indicating that the results are driven by the data rather than by the prior choice.

\section{Results}

\subsection{Constraints using the full dataset} 
\label{sec:full}
We estimate the VSH coefficients {up to degree $l=4$} using the full sample of 1,215,942 QSO-like sources with five-parameter astrometric solutions from Gaia DR3 \cite{2022A&A...667A.148G}. 
The fitted parameters include the three spheroidal $l=1$ coefficients (containing the peculiar acceleration of the Solar System through the first term in \eqref{ntildedrift}), as well as the three toroidal $l=1$ coefficients representing a global rotation. 
In addition, we estimate all ten quadrupole coefficients (five spheroidal and five toroidal) corresponding to the $l=2$ VSH modes, which capture possible {quadrupole} anisotropies in the proper motion field. All coefficients up to $l=2$ are estimated simultaneously as part of the full Bayesian analysis up to $l=4$ {(where the $l=3$ and $l=4$ coefficients are treated as nuisance parameters)}, as described in the previous section.

\begin{figure*}
    \includegraphics[width=0.49\textwidth]{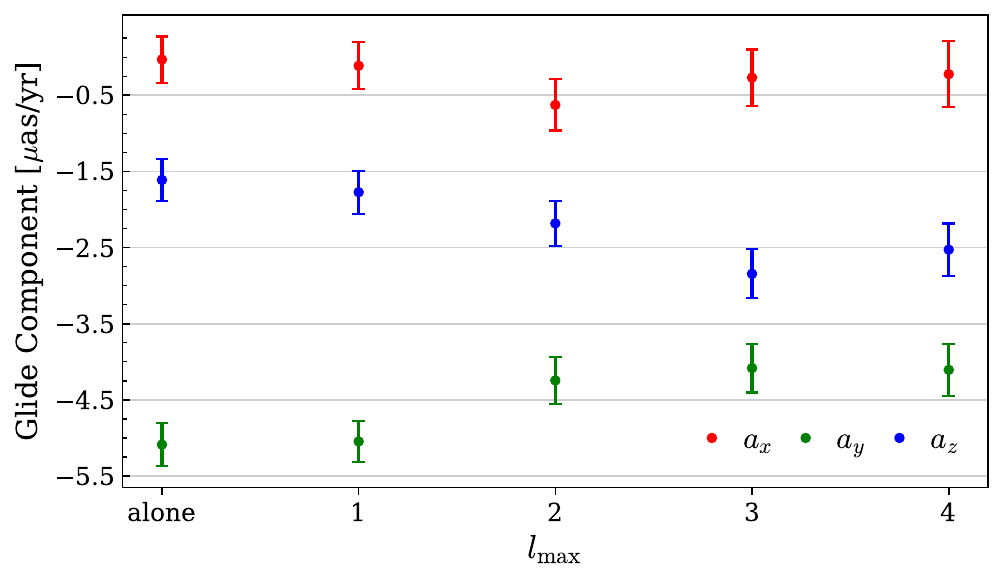}
    \includegraphics[width=0.49\textwidth]{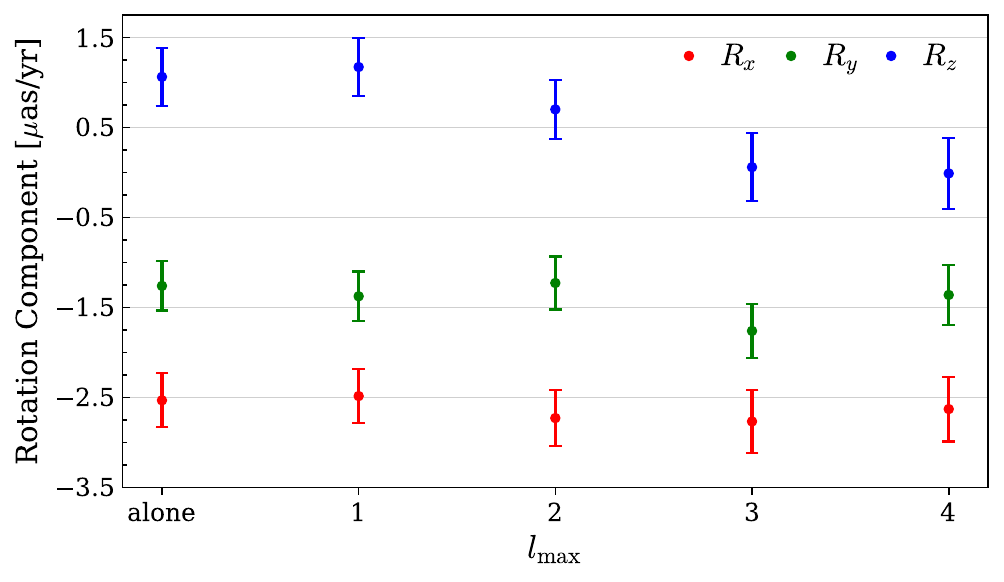}
    \caption{Estimated components of the {glide} vector (left) and the rotation vector (right) from the VSH fit. Error bars indicate the $1\sigma$ uncertainties. \sayy{alone} refers to the situation where all VSH coefficients apart from the glide vector (left) or the rotation vector (right) are set to zero. $l_{\max} = 1,2,3,4$ refers to the truncation of the VSH expansion at the given order.  
    }
    \label{fig:acceleration_rotation_components}
\end{figure*}

Figure~\ref{fig:acceleration_rotation_components} shows the fitted components of the acceleration and rotation vectors from the VSH analysis. All estimated coefficients up to $l=2$ are listed in Table~\ref{tab:vsh_results}. {We report the amplitudes of the glide (acceleration) vector, the rotation, and the total quadrupole,} and for the dipole, we also report its direction in equatorial coordinates.
{In Appendix~\ref{Appendix C}, we show the corresponding sky plots of the proper motion signal at large angular scales associated with the dipole and quadrupole components, as well as for the octupole components.}

{To assess the statistical significance of the quadrupole components, we compute the Bayes factors comparing the dipole model ($l_{\max}=1$) to the dipole+quadrupole model ($l_{\max}=2$). We see in Table~\ref{tab:vsh_results} that there is strong Bayesian support for a nonzero quadrupole with $\ln \mathcal{B} = 43$, which is also reflected in the high signal-to-noise ratios of the determined quadrupole amplitudes.}

\begin{figure*}
    \includegraphics[width=0.49\textwidth]{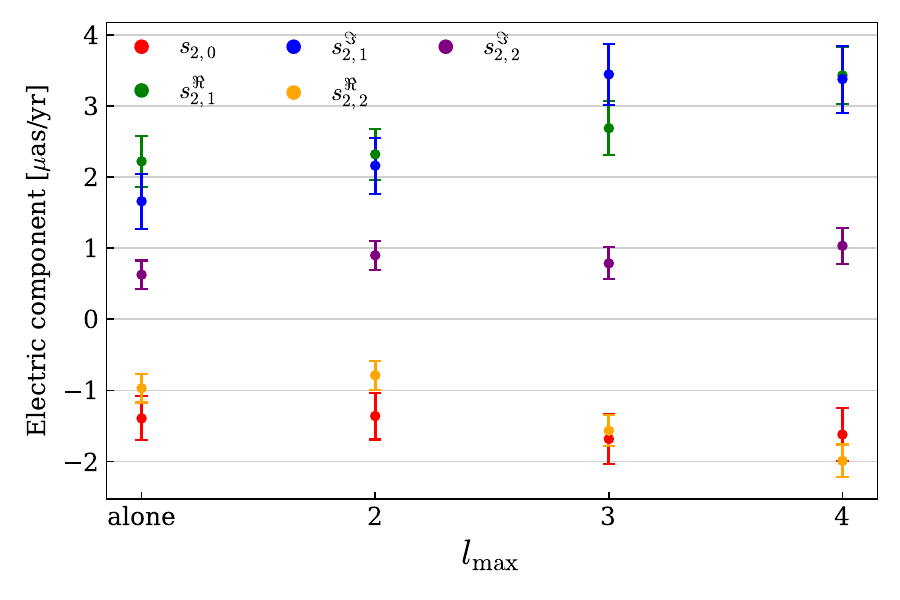}
    \includegraphics[width=0.49\textwidth]{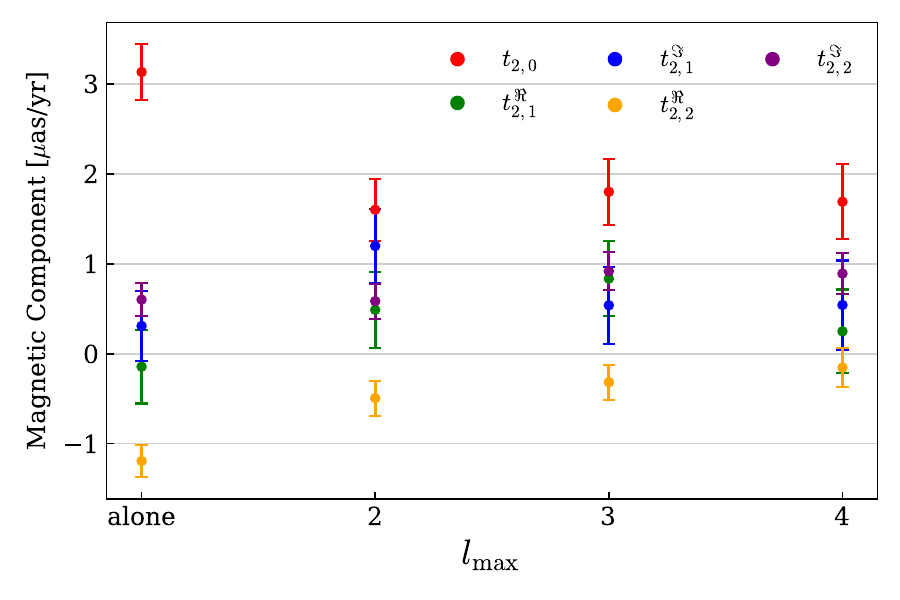}
    \caption{Estimated components of the electric quadrupole (left) and magnetic quadrupole (right) {from the VSH fit.} Error bars indicate the $1\sigma$ uncertainties. {\sayy{alone} refers to the setting where all VSH coefficients apart from the electric quadrupole (left) or the magnetic quadrupole (right) are set to zero. $l_{\max} = 2,3,4$ refers to the truncation of the VSH expansion at the given order.}} 
    \label{fig:quad_coeff}
\end{figure*}

The results for the glide vector agree with the results in \cite{2021A&A...649A...9G}, which is expected, because we are using the same dataset for our constraints. 
The amplitude of the rotation vector found is close to the result in eq.~12 of \cite{2022A&A...667A.148G}, and we assign the difference to being due to that \cite{2022A&A...667A.148G} includes the six-parameter\footnote{The six-parameter solutions include the pseudocolour as a parameter and are generally considered less accurate than the five-parameter solutions; see  \cite{2021A&A...649A...2L} for details.} sources in addition to the five-parameter sources that we consider. 
The non-vanishing rotation is due to the fact that there is a net-rotation of the sources in the CRF DR3 sources relative to the frame rotator sources that are used to calibrate the coordinate system of Gaia \cite{2022A&A...667A.148G}. 
{The amplitudes found for the electric and magnetic quadrupoles are in rough agreement with what is found in \cite{2022A&A...667A.148G} (see their section~3.4) and \cite{2021A&A...649A...2L} (see their section~5.6), where the presence of quadrupole and higher-order multipoles in the data are used as a measure of otherwise unquantified systematics. 
The observed quadrupole components in the proper motions may also be used to bound the gravitational wave background of gravitational waves with periods larger than the Gaia observation time \cite{gwenergydensity}.}

\begin{table}
%\small 
\setlength{\tabcolsep}{20pt}
\centering
\caption{Estimated dipole and quadrupole coefficients ($\mu$as/yr).}
\label{tab:vsh_results}
\begin{tabular}{lc}
%\toprule
\hline
\hline
\textbf{{Glide dipole}  } & \\
\( {a}_{x} \) & \( -0.22 \pm 0.43 \) \\
\( {a}_{y} \) & \( -4.11 \pm 0.34 \) \\
\( {a}_{z} \) & \( -2.53 \pm 0.34 \) \\
{$\abs{\bm{a}}$}  & \( \hspace{0.26cm} 4.86 \pm 0.34 \) \\
%\hline
%\textbf{Direction of the acceleration vector} & \\
{$\alpha$} %Right Ascension  
(\degree) & \( \hspace{0.27cm}266.9 \pm 6.0 \) \\
{$\delta$} %Declination 
(\degree) & \( -31.6 \pm 4.1 \) \\
\hline
\textbf{{Rotation dipole}} & \\
\( R_{x} \) & \( -2.63 \pm 0.36 \) \\
\( R_{y} \) & \( -1.36 \pm 0.33 \) \\
\( R_{z} \) & \( -0.01 \pm 0.40 \) \\
{$\abs{\bm{R}}$} & \( \hspace{0.26cm} 3.01 \pm 0.34 \) \\
\hline
\textbf{{Electric} Quadrupole} & \\
\( s_{2,0} \) & \( -1.63 \pm 0.38 \) \\
\( s_{2,1}^{\Re} \) & \( \hspace{0.26cm}3.42 \pm 0.41 \) \\
\( s_{2,1}^{\Im} \) & \( \hspace{0.26cm}3.36 \pm 0.46 \) \\
\( s_{2,2}^{\Re} \) & \( -1.99 \pm 0.23 \) \\
\( s_{2,2}^{\Im} \) & \( \hspace{0.26cm}1.03 \pm 0.26 \) \\
{$\abs{\bm{s}}$} & \( \hspace{0.26cm} 7.66 \pm 0.57 \) \\
\hline
\textbf{{Magnetic} Quadrupole} & \\
\( t_{2,0} \) & \( \hspace{0.26cm}1.69 \pm 0.42 \) \\
\( t_{2,1}^{\Re} \) & \( \hspace{0.26cm}0.29 \pm 0.46 \) \\
\( t_{2,1}^{\Im} \) & \( \hspace{0.26cm}0.53 \pm 0.50 \) \\
\( t_{2,2}^{\Re} \) & \( -0.15 \pm 0.21 \) \\
\( t_{2,2}^{\Im} \) & \( \hspace{0.26cm}0.90 \pm 0.23 \) \\
{$\abs{\bm{t}}$} & \( \hspace{0.26cm} 2.29 \pm 0.45 \) \\
\hline
\textbf{Quadrupole Amplitude} & \(\hspace{0.26cm} 7.99 \pm 0.56 \) \\
\hline
\textbf{Bayes Factor}\\
$\ln \mathcal{B}$ & \(\hspace{0.26cm} 43 \)\\ %43.35
%\toprule
\hline
\hline
\end{tabular}
\end{table}

\subsection{Redshift dependence}
\label{sec:zdep}
\begin{figure*}
    \includegraphics[width=0.49\textwidth]{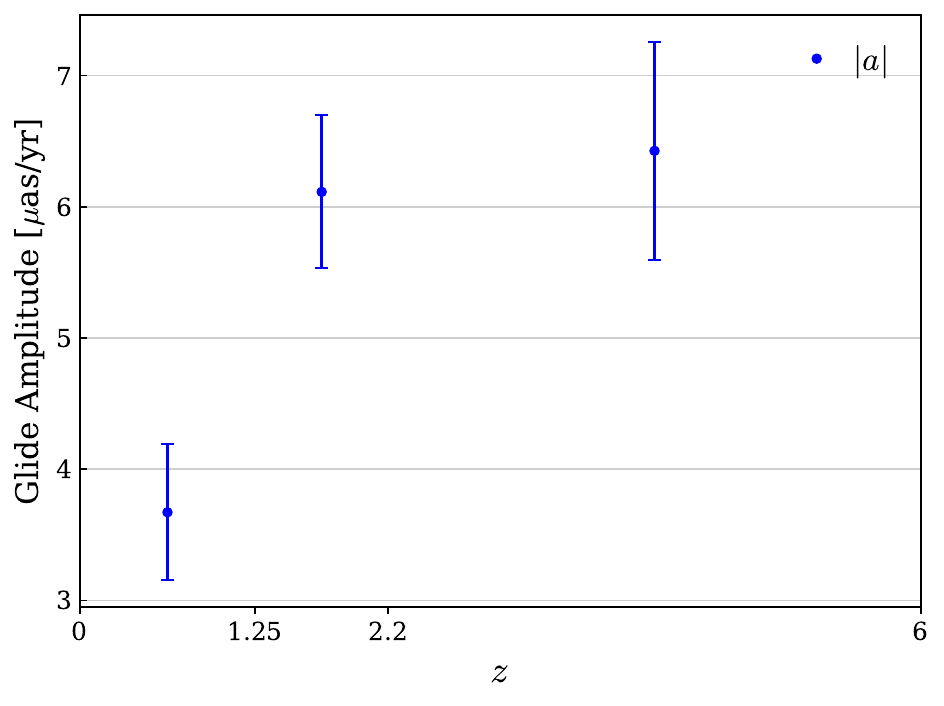}
    \includegraphics[width=0.49\textwidth]{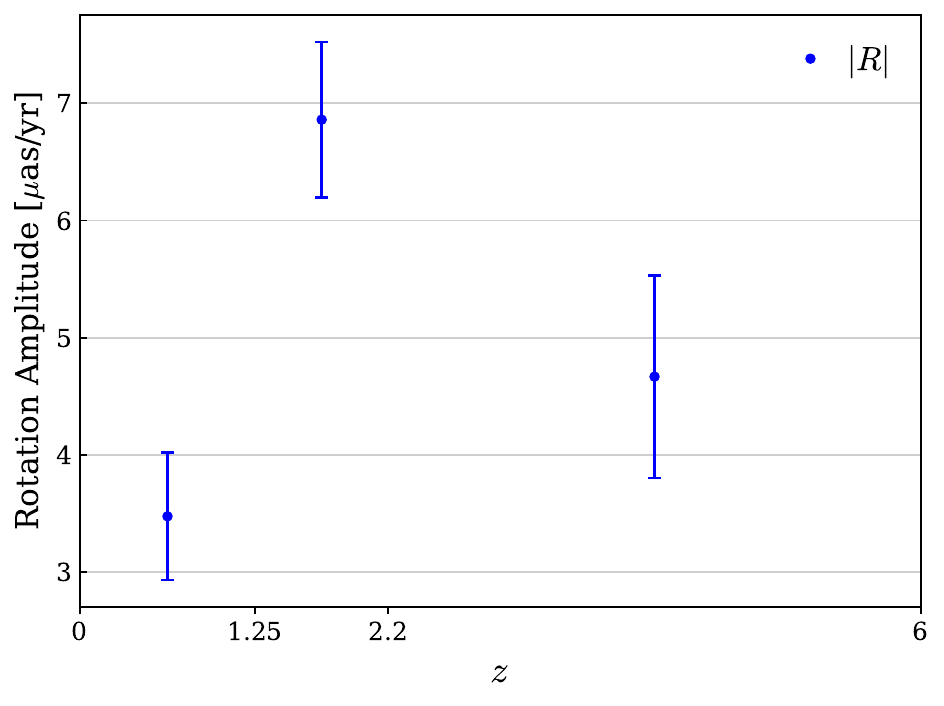}
    \includegraphics[width=0.49\textwidth]{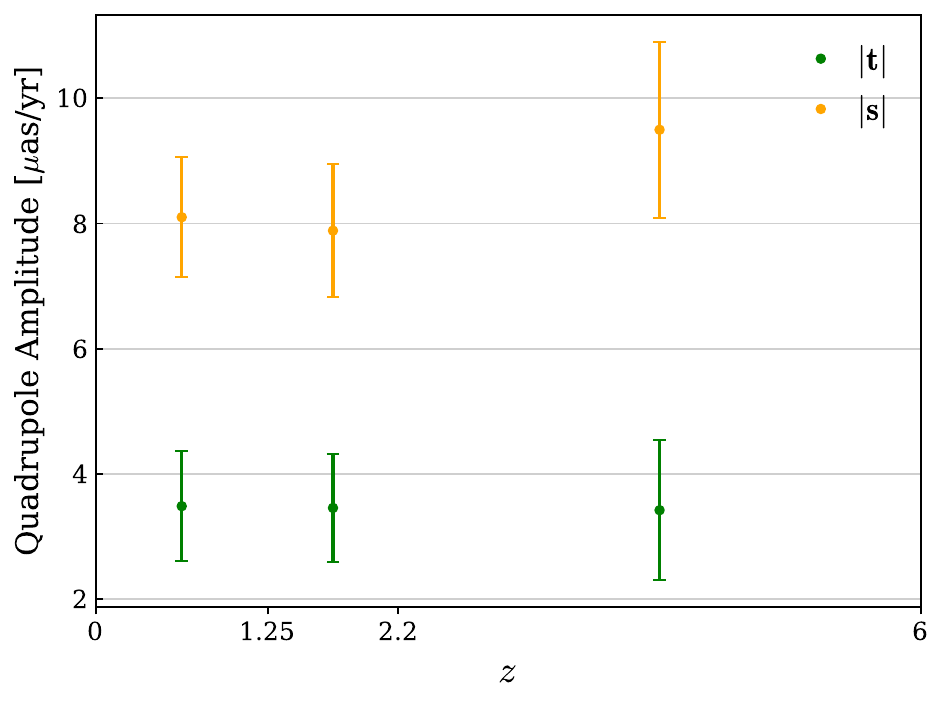}
    \includegraphics[width=0.49\textwidth]{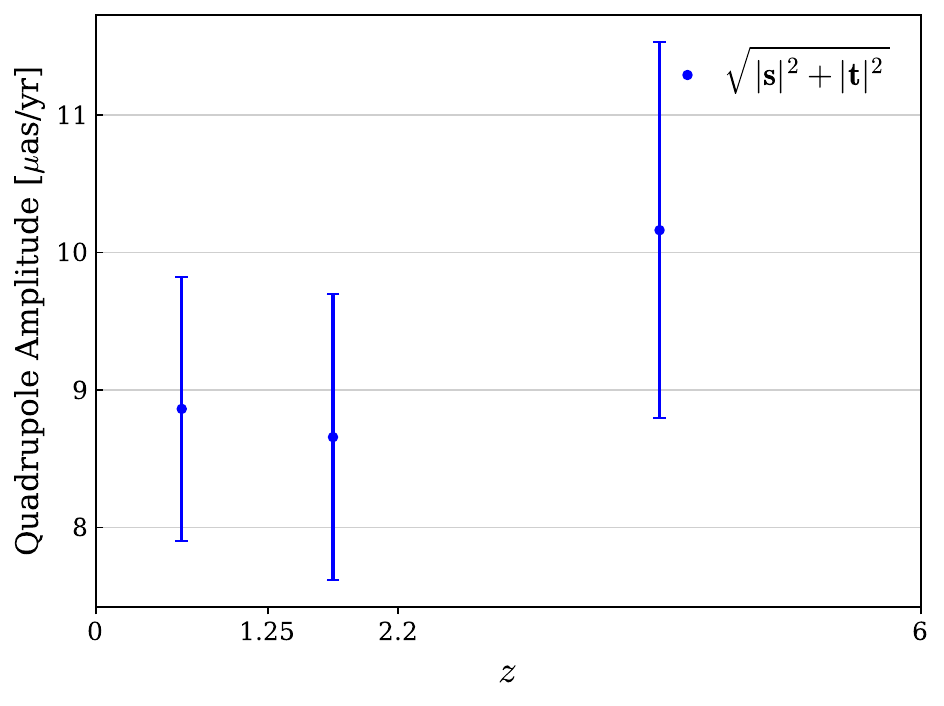}
    \caption{Estimated dipole and quadrupole components as a function of redshift. The dataset is divided to ensure approximately equal numbers of sources per bin. Bin 1: 0–1.25 (395651 sources), Bin 2: 1.25-2.2 (392000 sources), Bin 3: 2.2-6 (378734 sources). 
    }
    \label{fig:redshift_dependence}
\end{figure*}

To examine the potential redshift dependence of the dipole and quadrupole components, we divide the dataset into three redshift bins, each containing approximately the same number of sources. This binning ensures comparable statistical weight across subsets. %We restrict the analysis to quasars with redshift $z > 0.023$ in order to minimize contamination from sources within {an approximate radius of homogeneity of the cosmological matter distribution.} 
We exclude sources for which no redshift information is available and sources with $z> 6$. 
After applying these cuts {(which turn out to make very little difference for our results)}, the resulting sample consists of {1,166,385} quasars. In each bin, we estimate the relevant VSH coefficients independently using the same Bayesian procedure applied to the full sample. The results are shown in Fig.~\ref{fig:redshift_dependence} 

{The glide vector amplitude increases significantly with redshift, reaching nearly double the value in the highest bin compared to the lowest. 
The visible appearance of the trend is mildly dependent on the binning, but there appears to be a robust jump from low redshifts to high redshifts with a significance in the range $2-3\sigma$.} 
{While this could indicate a genuine cosmological signal, which would be interesting in that it would conflict with the usual interpretation of the signal as solely being due to acceleration of the Solar System, it is important to consider potential sources of systematic error}. 
{Since the calibration models used for subsamples within Gaia are correlated with the magnitudes of the sources \cite{2020A&A...633A...1L}, which are again correlated with the redshifts {(see Fig.~\ref{fig:z_distribution})}, this could potentially result in systematics for the dipole with redshift (although this is usually discussed in relation to the toroidal part of the signal, as commented on below). } 

{The amplitude of the rotation vector varies between bins with significance around $3\sigma$ or higher for the largest jump. } 
This effect could be due to well-known differences between bright and faint sources in the CRF3 dataset, which have been suggested to be due to limitations of the calibration of the astrometric instruments \cite{2020A&A...633A...1L,2021A&A...649A...2L,2025arXiv250618758Z}. 
{We find no significant trend in the electric or magnetic quadrupole amplitudes.}  

\begin{figure}
    \includegraphics[width=0.49\textwidth]{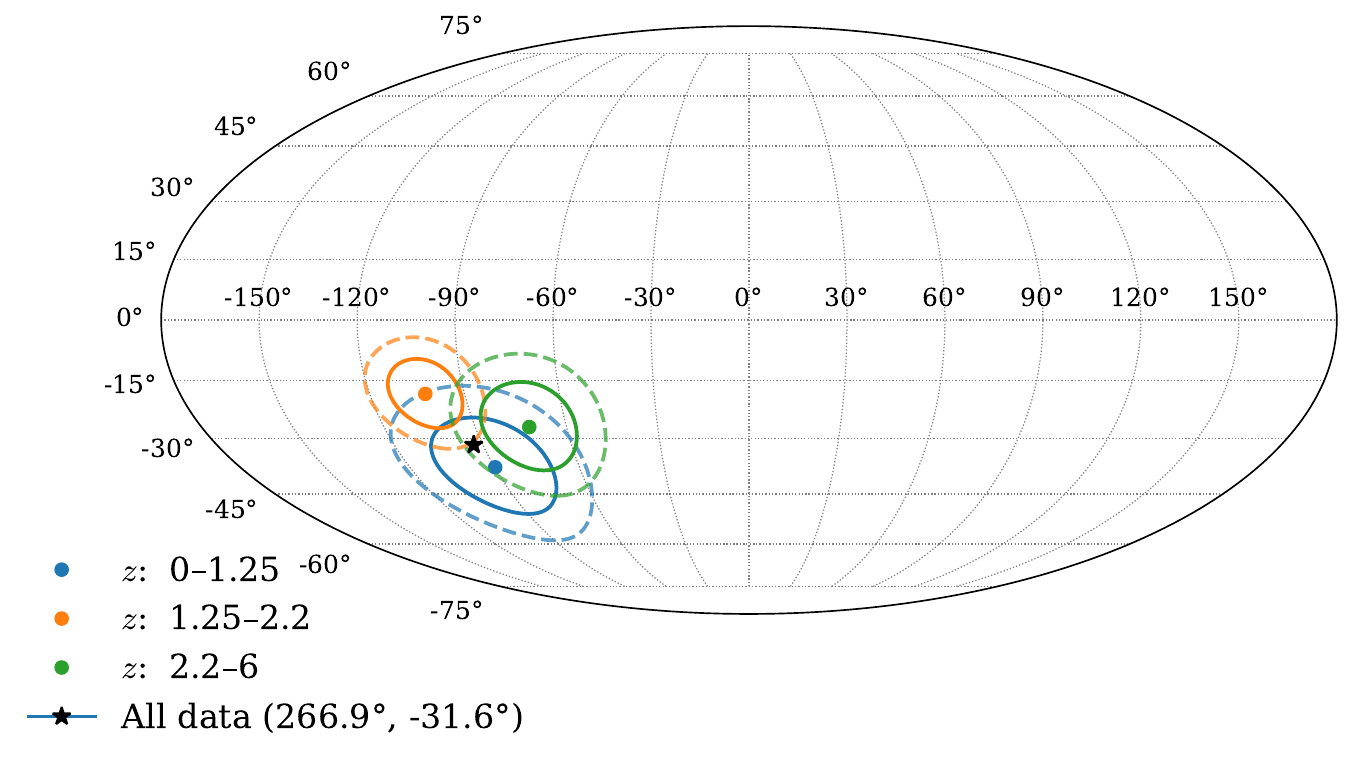}
    \caption{Glide vector direction with 1-sigma (solid) and 2-sigma (dashed) ellipses for each redshift bin. The black star is the all–data $l_{\max}=4$ fit. }
    \label{fig:glide_direction}
\end{figure}

{The direction vectors of the dipole components appear more robust across redshifts than their amplitudes. 
The constraints on the direction of the glide vector are shown in Fig.~\ref{fig:glide_direction} for the same redshift bins as for the amplitude analysis. The estimated directions are internally consistent with each other at the 2-sigma level.  }  

\begin{figure}
    \includegraphics[width=0.49\textwidth]{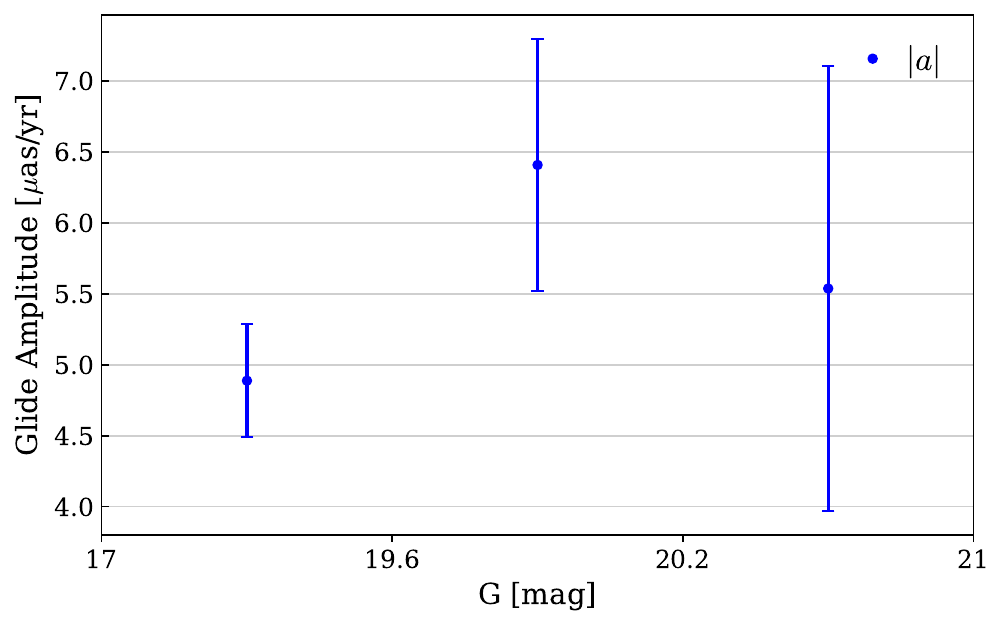}
    \includegraphics[width=0.49\textwidth]{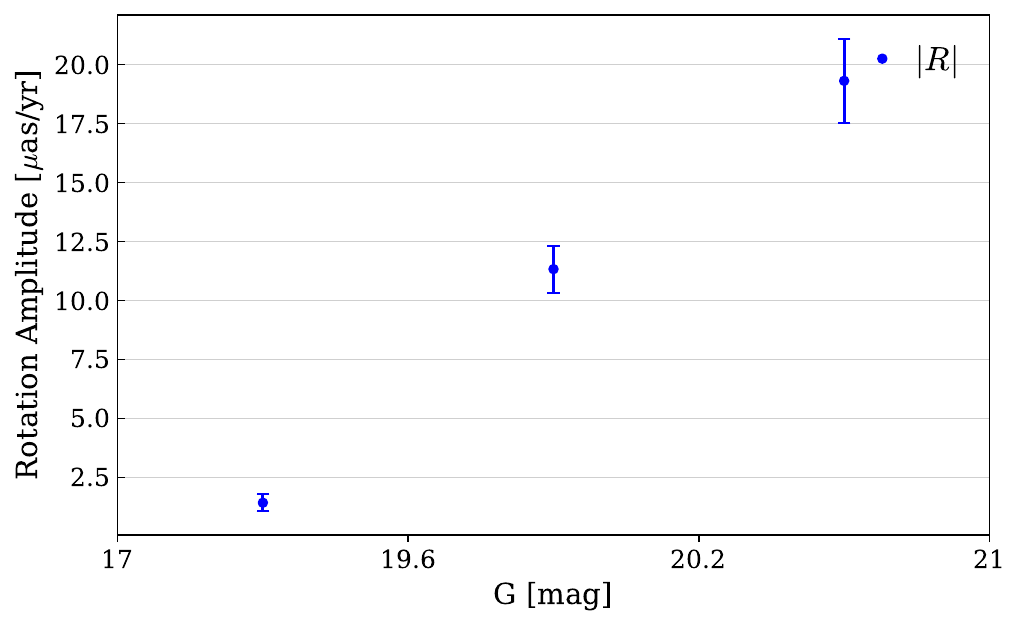}
    \caption{Estimated {glide vector (top) and rotation vector (bottom) amplitudes} as a function of the G-band magnitude{, G}. The dataset is divided to ensure approximately
equal numbers of sources per bin. Bin 1: 17-19.6 mag (412384 sources), Bin 2: 19.6-20.2 mag (397928 sources), Bin 3: 20.2-21 mag (401812 sources). 
}
    \label{fig:magnitude_dependence}
\end{figure}

{Since instrumentation calibration errors in Gaia are expected to correlate with magnitude and should thus to some degree also correlate with redshift, we examine the magnitude dependence directly in Fig.~\ref{fig:magnitude_dependence}. There is a clear dependence on the rotation vector with {the G-band} magnitude {(much stronger than the redshift dependence in Fig.~\ref{fig:redshift_dependence})}, thus indicating that the redshift dependence of the rotation vector is indeed likely due to systematics in calibration {. However, we do not see any significant dependence of the glide vector on magnitude, and  therefore it appears that the trend of the glide vector in redshift cannot be assigned to instrumental calibration errors of the type expected for the rotation vector.}    } 

We have also examined whether the dependence on redshift could be a spurious effect due to leakage of higher-order multipoles of order $l>4$. Although our convergence results for the full CRF3 sample in  Fig.~\ref{fig:acceleration_rotation_components} already indicate that convergence has been reached, we perform a separate analysis of the redshift-binned sub-samples in Appendix~\ref{Appendix B}. The results of the appendix indicate that convergence has indeed been reached for $l_{\max} =4$, and confirm the robustness of the results presented in this section. 

One may also speculate that the errors in the identification of redshifts\footnote{Gaia’s spectro-photometric classification for quasars is known to be less reliable in certain redshift ranges, such as $0.9<z<1.3$, where the $Mg_{II}$ line is often the only clearly visible emission feature in the spectra, and around $z\approx 2$, where the $C_{IV}$ line can be misidentified as Ly$\alpha$ \cite{Delchambre:2022ugo}.} of the sources could lead to systematics in the redshift-binned analysis. 
{We however believe that this is unlikely to cause any spurious redshift dependencies in our analysis, because (i) even large systematic errors in the redshift determinations should not alter a constant-in-redshift signal; (ii) the vast majority of objects are identified with an error in redshift $<0.1$ \cite{Delchambre:2022ugo} which is much less than the sizes of our redshift bins.   }

Thus, we conclude that the trend of the glide vector in redshift observed in Fig.~\ref{fig:redshift_dependence} is due to (i) astrophysical systematic errors that seem to not be {attributed} to magnitude-dependent calibration errors; (ii) a real cosmological trend in the data; or (iii) a statistical fluctuation at the $2-3\sigma$ level.

\section{Discussion} 
\label{sec:discussion} 
The redshift dependence found in our results hint towards a factor of $\sim 2$ discrepancy between determinations of the acceleration at moderate redshifts $0 < z <1.25$ and at high redshifts $z > 1.25$. 
A previous estimate of redshift dependence in complementary very long baseline interferometry (VLBI) data showed consistency with a constant dipole signal \cite{2011A&A...529A..91T}, however their results are not in conflict with ours due to the {wide error bars of their analysis}. These results thus do not hint towards whether the trend is induced by systematics or is a genuine cosmological signal. 
{More recent investigations by \citet{Makarov:2025vtb}, using the CRF dataset of Gaia DR3, hinted at redshift dependence of the glide vector, with preference from a growth of glide amplitude with redshift; cf. their Table~2. Their statistical evidence for the growth appear a bit weaker than what is found in this paper, which could be due to the difference in choice of redshift bins as well as the difference in the CRF dataset, where \citet{Makarov:2025vtb} includes the objects with six-parameter solutions.} 

If the source of the redshift dependence of the glide vector is an unquantified systematic error, for instance induced by the calibration of astrometric instruments, this warrants a careful examination of its origin in order to ensure an accurate measurement of the Solar System acceleration.  
We have examined the dependence of the glide vector signal with magnitude in order to assess this hypothesis, but we find consistency with a constant signal across magnitudes as was also previously found \cite{2021A&A...649A...9G}, indicating that the redshift dependence found in our work does not simply arise through a problem with magnitude-correlated systematics. 

If the redshift dependence rather arises from the cosmology itself, this opens up a plethora of possibilities. 
We remark that simple order-of-magnitude estimates seem to exclude an easy interpretation in terms of motion effects within the $\Lambda$CDM model: Drift rates of the order of $1 \,
\mu\textrm{as}\,\textrm{yr}^{-1}$ correspond to transverse velocities of $0.05\,c$ and $0.13\,c$ for objects at $z=0.1$ and $z=6$ respectively, assuming a standard FLRW metric with Planck values for the cosmological parameters. This would be the scale of transverse velocities needed to account for the spheroidal dipole variations with redshift of the order of $1 \,
\mu\textrm{as}\,\textrm{yr}^{-1}$. 
A purely kinematical explanation of the results would thus require huge velocities at large distances. 

The quadrupole components are most often assigned to unquantified systematic errors \cite{2021A&A...649A...2L,2022A&A...667A.148G} or as possible probes of long-wavelength gravitational waves \cite{gwenergydensity,Darling:2018hmc} or large-scale anisotropic expansion of space as, for instance, in a Bianchi space-time \cite{Marcori:2018cwn}. 
The quadrupole components found in this analysis are of the order $8\  \mu\textrm{as}\,\textrm{yr}^{-1}$, and applying the above order-of-magnitude calculation again suggests that a kinematic interpretation within the $\Lambda$CDM framework is excluded. 
We do not observe any significant trends of the electric or magnetic quadrupole components or amplitudes with redshift. 

The rotation vector has a $\sim 3 \sigma$ significant dependence on redshift bin in our analysis. The variation of the rotation vector between different subsamples of Gaia DR3 has been noted before, and has been assigned to differences in astrometric calibration models at different magnitudes  \cite{2020A&A...633A...1L,2021A&A...649A...2L}. Our work supports this conclusion in the sense that we observe even stronger trends of the rotation vector with magnitude than we do for redshift, and it is plausible that the redshift dependence that we do see simply enters through the correlation of the measured redshifts with magnitude. 
The rotation vector can constrain models with large-scale vorticity in cosmology \cite{2012ApJ...755...58N,Amendola:2013bga}, and it is therefore of fundamental interest to quantify and reduce the systematic errors associated with calibration. 

While Gaia {data} has been {widely} explored for its astrophysical constraints, such as the ability to map the Milky Way's structure, evolution, and stellar streams, there has been relatively little focus on the cosmological implications. 
We want to highlight the great promise of Gaia data for answering questions of fundamental importance in cosmology; see also the review papers \cite{Quercellini:2010zr,Darling:2018gxk}. 
Proper motion estimates of QSOs are thus important for deepening our understanding of the cosmological rest frame, for tests of the cosmological principle \cite{2022ApJ...927L...4M}, null tests of the $\Lambda$CDM model and tests of alternative cosmological scenarios \cite{Krasinski:2010rc,Quercellini:2010zr,Amendola:2013bga}, for constraining the gravitational wave background \cite{gwenergydensity,Darling:2018hmc}, and for mapping the metric of our cosmological neighborhood with cosmography methods \cite{Heinesen:2024npe}.

One of the current tensions in cosmology of reported strong statistical significance arises from the Ellis-Baldwin measurement of our velocity relative to the baryonic rest frame through number counts of distant sources \cite{1984MNRAS.206..377E}, which has been reported to be in strong disagreement with our velocity relative to the cosmic microwave background \cite{Secrest:2025nbt}. 

This further highlights the importance of inferring accurate complementary constraints on the cosmic restframe from Gaia observations. 
It would be interesting to be able to do a consistency test of our peculiar acceleration with respect to the distant QSOs (assumed to represent the baryonic restframe) as measured by Gaia and our peculiar acceleration with respect to the cosmic microwave background radiation in the spirit of the Ellis-Baldwin test. 
However, we do not have a precise measurement of our peculiar acceleration relative to the cosmic microwave background radiation, which would require a measurement of the drift of the cosmic microwave background temperature with a relative precision of $\sim 10^{-10}$, which can be compared with the accuracy of the Planck measurement of $10^{-5}$. 
If an estimate of our peculiar acceleration relative to the background radiation field was made possible, it would be a very interesting complementary rest frame test. 
Other valuable information from the proper motion signal is the velocity of the Solar System itself, which enters in the last term of \eqref{ntildedrift}, and which could give an important independent measurement of our velocity relative to the baryonic matter in our Universe. Since this term is heavily suppressed by the inverse distance factor multiplying the velocity, the current errors of Gaia on the estimates of the glide vector must be reduced by at least an order of magnitude to constrain realistic velocities of a few $100$ \,km/s \cite{Paine:2019vep}.

Future Gaia data releases have the potential to investigate the suggestive results found in this paper, and to go further in extracting important cosmological information from the drift effects. The accuracy of Solar System acceleration measurements has been estimated to improve by a factor of $\sim 10$ with data release 4 through a combination of more observation time for each source, additional sources, and better control of systematic errors \cite{2021A&A...649A...9G} (see their section~8). 
A future Gaia Near Infra-Red telescope may have the potential to reduce statistical errors by a factor of $\sim 14$ when combined with the current Gaia datasets \cite{2021ExA....51..783H}, and could also be important for reducing and better estimating systematic errors because of the independent systematic errors relative to the Gaia mission.

\section{Conclusion} 
\label{sec:conclusion} 
The glide component of the proper motions of QSOs is usually interpreted as quantifying the acceleration of the Solar System with respect to the global baryonic restframe defined by distant QSOs \cite{2021A&A...649A...9G}. 

{We have examined the proper motions of the CRF sources contained in Gaia DR3 in order to constrain the lowest-order vector spherical harmonic contributions to the proper-motion field across the sky. 
We examined the redshift dependence of the dipole and quadrupole signals in order to test the null hypothesis that the signal is solely explained by the local acceleration of the Solar System. Our results for the glide vector are in mild tension with this null hypothesis, with a constant glide vector across redshifts being ruled out at the $2-3\sigma$ level. 
Although this result could point to new cosmological physics, there are possible systematic errors that may induce redshift dependence of the proper motion signal. Our analysis indicates that magnitude-dependent calibration errors are not likely to explain the full trend, but this should be examined in more detail as well as other possible systematics. 
The presence of non-zero quadrupole components (that appear stable across redshifts) of the same order as the dipole amplitude also hints at systematic errors and/or new physics in the data to be understood. 

We have discussed possible sources of systematic errors as well as potential new cosmological phenomena that could give rise to the signatures seen in our analysis. 
Upcoming data releases with more constraining power are needed to either confirm our findings with more significance or to identify them as due to a statistical outlier or systematic errors in the data.  

In our analysis, we confirm that the glide vector of the full CRF3 sample is in agreement with the expected acceleration of the Solar System based on estimates that can be made from gravitational potentials of the Milky Way and large clusters of galaxies in our cosmic neighborhood \cite{2021A&A...649A...9G}. 
However, our analysis also reveals a significant trend in the signal with redshift, a finding that is approximately five times larger than the statistical error of the full sample -- a result that warrants further consideration. 

}

\vspace{6pt} 
%%%%%%%%%%%%%%%%%%%%%%%%%%%%%%%%%%%%%%%%%%%%%%%%%%%%%%%%%%%%%%%%%%%
\section*{Acknowledgements}
{During the final state of the writing of our paper, we discovered that another paper which also examines redshift dependence of the low-order multipoles in Gaia data, has been carried out in parallel to ours by \citet{Makarov:2025vtb}; see Section~\ref{sec:discussion} for comments on the results of this paper in relation to ours.    } 
We would like to thank Sebastian von Hausegger and Subir Sarkar for discussions.  
AH is supported by the Carlsberg Foundation. 
This project has received support from the Villum Foundation (Project No. 13164, PI: I. Tamborra), the Danmarks Frie Forskningsfond (Project No. 8049-00038B, PI: I. Tamborra).

\bibliographystyle{aa}
\bibliography{refs}

\clearpage

\appendix

\onecolumn
\section{Vector spherical harmonics for $l\le4$}
\label{Appendix A}

{In Table~\ref{tab:vsh4_s} and~\ref{tab:vsh4_t} we list the spheroidal and toroidal VSH basis functions up to and $l=4$}
   
\begin{table*}
\centering
\caption{Spheroidal harmonics up to $l =4$.}
\label{tab:vsh4_s}
\setlength{\tabcolsep}{25pt}
\begin{tabular}{lcc}
\hline
{\bf Harmonics} & {\bf $\mu_{\alpha\ast}$} & {\bf $\mu_{\delta}$} \\
\hline
$\qquad S_{10}$            & ---                                         & $\cos\delta$                                      \\[1ex]
$\qquad S_{11}^{\rm \Re}$   & $\sin\alpha$                                & $\sin\delta\cos\alpha$                            \\[1ex]
$\qquad S_{11}^{\rm \Im}$   & $-\cos\alpha$                               & $\sin\delta\sin\alpha$                            \\[1ex]\hline
$\qquad S_{20}$            & ---                                         & $\sin2\delta$                                     \\[1ex]
$\qquad S_{21}^{\rm \Re}$   & $\sin\delta\sin\alpha$                      & $-\cos2\delta\cos\alpha$                          \\[1ex]
$\qquad S_{21}^{\rm \Im}$   & $-\sin\delta\cos\alpha$                     & $-\cos2\delta\sin\alpha$                          \\[1ex]
$\qquad S_{22}^{\rm \Re}$   & $-2\cos\delta\sin2\alpha$                   & $-\sin2\delta\cos2\alpha$                         \\[1ex]
$\qquad S_{22}^{\rm \Im}$   & $2\cos\delta\cos2\alpha$                    & $-\sin2\delta\sin2\alpha$                         \\[1ex]\hline
$\qquad S_{30}$            & ---                                         & $\cos\delta(5\sin^2\delta - 1)$                   \\[1ex]
$\qquad S_{31}^{\rm \Re}$   & $(5\sin^2\delta - 1)\sin\alpha$             & $\sin\delta(15\sin^2\delta - 11)\cos\alpha$       \\[1ex]
$\qquad S_{31}^{\rm \Im}$   & $-(5\sin^2\delta - 1)\cos\alpha$            & $\sin\delta(15\sin^2\delta - 11)\sin\alpha$       \\[1ex]
$\qquad S_{32}^{\rm \Re}$   & $-\sin2\delta\sin2\alpha$                   & $-\cos\delta(3\sin^2\delta - 1)\cos2\alpha$       \\[1ex]
$\qquad S_{32}^{\rm \Im}$   & $\sin2\delta\cos2\alpha$                    & $-\cos\delta(3\sin^2\delta - 1)\sin2\alpha$       \\[1ex]
$\qquad S_{33}^{\rm \Re}$   & $\cos^2\delta\sin3\alpha$                   & $\cos^2\delta\sin\delta\cos3\alpha$               \\[1ex]
$\qquad S_{33}^{\rm \Im}$   & $-\cos^2\delta\cos3\alpha$                  & $\cos^2\delta\sin\delta\sin3\alpha$               \\[1ex]\hline
$\qquad S_{40}$   & ---                          & $\sin2\delta(7\sin^2\!\delta-3)$                    \\[1ex]
$\qquad S_{41}^{\rm \Re}$   & $\sin\delta(7\sin^2\!\delta-3)\sin\alpha$                         & $(28\sin^4\!\delta-27\sin^2\!\delta+3)\cos\alpha$                   \\[1ex]
$\qquad S_{41}^{\rm \Im}$            & $-\sin\delta(7\sin^2\!\delta-3)\cos\alpha$                     & $(28\sin^4\!\delta-27\sin^2\!\delta+3)\sin\alpha$                                         \\[1ex]
$\qquad S_{42}^{\rm \Re}$   & $-\cos\delta(7\sin^2\!\delta-1)\sin2\alpha$         & $-\sin2\delta(7\sin^2\!\delta-4)\cos2\alpha$              \\[1ex]
$\qquad S_{42}^{\rm \Im}$   & $\cos\delta(7\sin^2\!\delta-1)\cos2\alpha$         & $-\sin2\delta(7\sin^2\!\delta-4)\sin2\alpha$               \\[1ex]
$\qquad S_{43}^{\rm \Re}$   & $3\cos^2\!\delta\sin\delta\sin3\alpha$         & $\cos^2\!\delta(4\sin^2\!\delta-1)\cos3\alpha$                    \\[1ex]
$\qquad S_{43}^{\rm \Im}$   & $-3\cos^2\!\delta\sin\delta\cos3\alpha$         & $\cos^2\!\delta(4\sin^2\!\delta-1)\sin3\alpha$                   \\[1ex]
$\qquad S_{44}^{\rm \Re}$   & $-\cos^3\!\delta\sin4\alpha$               & $-\cos^3\!\delta\sin\delta\cos4\alpha$                  \\[1ex]
$\qquad S_{44}^{\rm \Im}$   & $\cos^3\!\delta\cos4\alpha$               & $-\cos^3\!\delta\sin\delta\sin4\alpha$                   \\[1ex]
\hline
\end{tabular}
\end{table*}

\begin{table*}
\centering
\caption{Toroidal harmonics up to $l = 4$. }
\label{tab:vsh4_t}
\setlength{\tabcolsep}{25pt}
\begin{tabular}{lcc}
\hline
{\bf Harmonics} & {\bf $\mu_{\alpha\ast}$} & {\bf $\mu_{\delta}$}\\
\hline
$\qquad T_{10}$            & $\cos\delta$                                      & ---                                         \\[1ex]
$\qquad T_{11}^{\rm \Re}$   & $\sin\delta\cos\alpha$                            & $-\sin\alpha$                               \\[1ex]
$\qquad T_{11}^{\rm \Im}$   & $\sin\delta\sin\alpha$                            & $\cos\alpha$                                \\[1ex]\hline
$\qquad T_{20}$            & $\sin2\delta$                                     & ---                                         \\[1ex]
$\qquad T_{21}^{\rm \Re}$   & $-\cos2\delta\cos\alpha$                          & $-\sin\delta\sin\alpha$                     \\[1ex]
$\qquad T_{21}^{\rm \Im}$   & $-\cos2\delta\sin\alpha$                          & $\sin\delta\cos\alpha$                      \\[1ex]
$\qquad T_{22}^{\rm \Re}$   & $-\sin2\delta\cos2\alpha$                         & $2\cos\delta\sin2\alpha$                    \\[1ex]
$\qquad T_{22}^{\rm \Im}$   & $-\sin2\delta\sin2\alpha$                         & $-2\cos\delta\cos2\alpha$                   \\[1ex]\hline
$\qquad T_{30}$            & $\cos\delta(5\sin^2\delta-1)$                     & ---                                         \\[1ex]
$\qquad T_{31}^{\rm \Re}$   & $\sin\delta(15\sin^2\delta-11)\cos\alpha$         & $-(5\sin^2\delta-1)\sin\alpha$              \\[1ex]
$\qquad T_{31}^{\rm \Im}$   & $\sin\delta(15\sin^2\delta-11)\sin\alpha$         & $(5\sin^2\delta-1)\cos\alpha$               \\[1ex]
$\qquad T_{32}^{\rm \Re}$   & $-\cos\delta(3\sin^2\delta-1)\cos2\alpha$         & $\sin2\delta\sin2\alpha$                    \\[1ex]
$\qquad T_{32}^{\rm \Im}$   & $-\cos\delta(3\sin^2\delta-1)\sin2\alpha$         & $-\sin2\delta\cos2\alpha$                   \\[1ex]
$\qquad T_{33}^{\rm \Re}$   & $\cos^2\delta\sin\delta\cos3\alpha$               & $-\cos^2\delta\sin3\alpha$                  \\[1ex]
$\qquad T_{33}^{\rm \Im}$   & $\cos^2\delta\sin\delta\sin3\alpha$               & $\cos^2\delta\cos3\alpha$                   \\[1ex]\hline
$\qquad T_{40}$   & $\sin2\delta(7\sin^2\!\delta-3)$                         & ---                     \\[1ex]
$\qquad T_{41}^{\rm \Re}$   & $(28\sin^4\!\delta-27\sin^2\!\delta+3)\cos\alpha$                         & $-\sin\delta(7\sin^2\!\delta-3)\sin\alpha$                   \\[1ex]
$\qquad T_{41}^{\rm \Im}$            & $(28\sin^4\!\delta-27\sin^2\!\delta+3)\sin\alpha$                     & $\sin\delta(7\sin^2\!\delta-3)\cos\alpha$                                         \\[1ex]
$\qquad T_{42}^{\rm \Re}$   & $-\sin2\delta(7\sin^2\!\delta-4)\cos2\alpha$         & $\cos\delta(7\sin^2\!\delta-1)\sin2\alpha$              \\[1ex]
$\qquad T_{42}^{\rm \Im}$   & $-\sin2\delta(7\sin^2\!\delta-4)\sin2\alpha$         & $-\cos\delta(7\sin^2\!\delta-1)\cos2\alpha$               \\[1ex]
$\qquad T_{43}^{\rm \Re}$   & $\cos^2\!\delta(4\sin^2\!\delta-1)\cos3\alpha$         & $-3\cos^2\!\delta\sin\delta\sin3\alpha$                    \\[1ex]
$\qquad T_{43}^{\rm \Im}$   & $\cos^2\!\delta(4\sin^2\!\delta-1)\sin3\alpha$         & $3\cos^2\!\delta\sin\delta\cos3\alpha$                   \\[1ex]
$\qquad T_{44}^{\rm \Re}$   & $-\cos^3\!\delta\sin\delta\cos4\alpha$               & $\cos^3\!\delta\sin4\alpha$                  \\[1ex]
$\qquad T_{44}^{\rm \Im}$   & $-\cos^3\!\delta\sin\delta\sin4\alpha$               & $-\cos^3\!\delta\cos4\alpha$                   \\[1ex]
\hline
\end{tabular}
\end{table*}

\twocolumn

\section{Molleweide projection of proper motions}
\label{Appendix C}

\begin{figure}
    \includegraphics[width=0.49\textwidth]{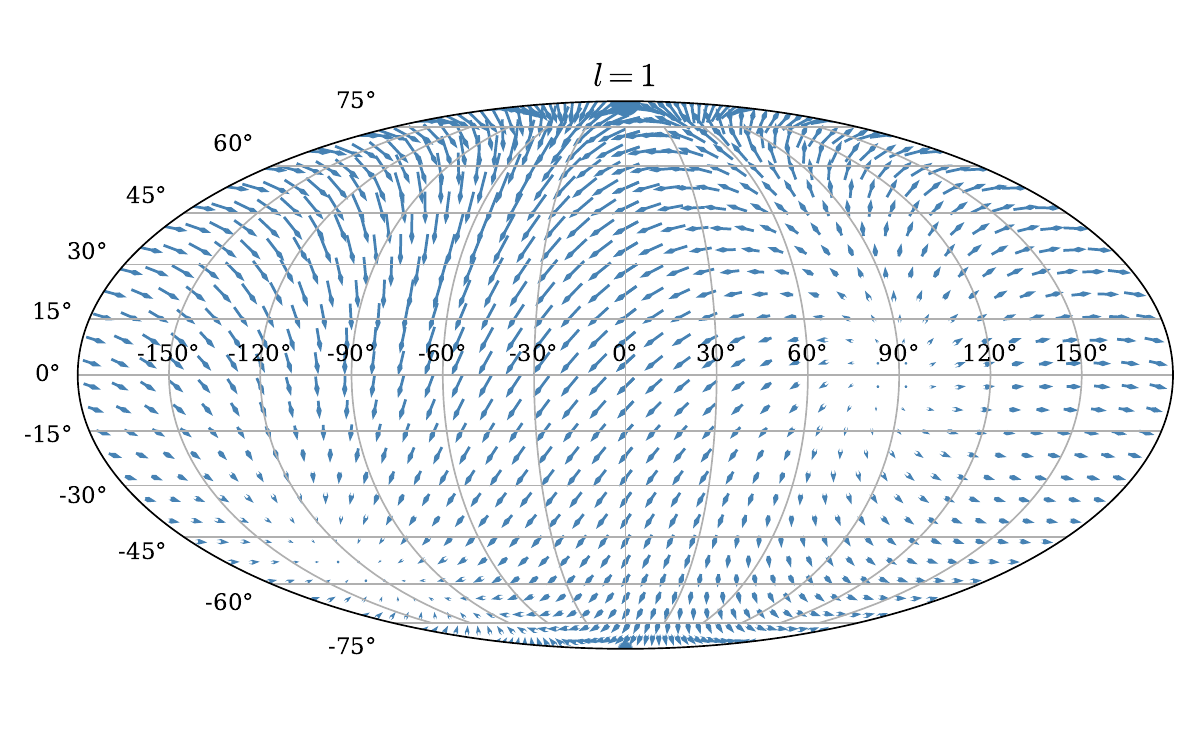}
    \includegraphics[width=0.49\textwidth]{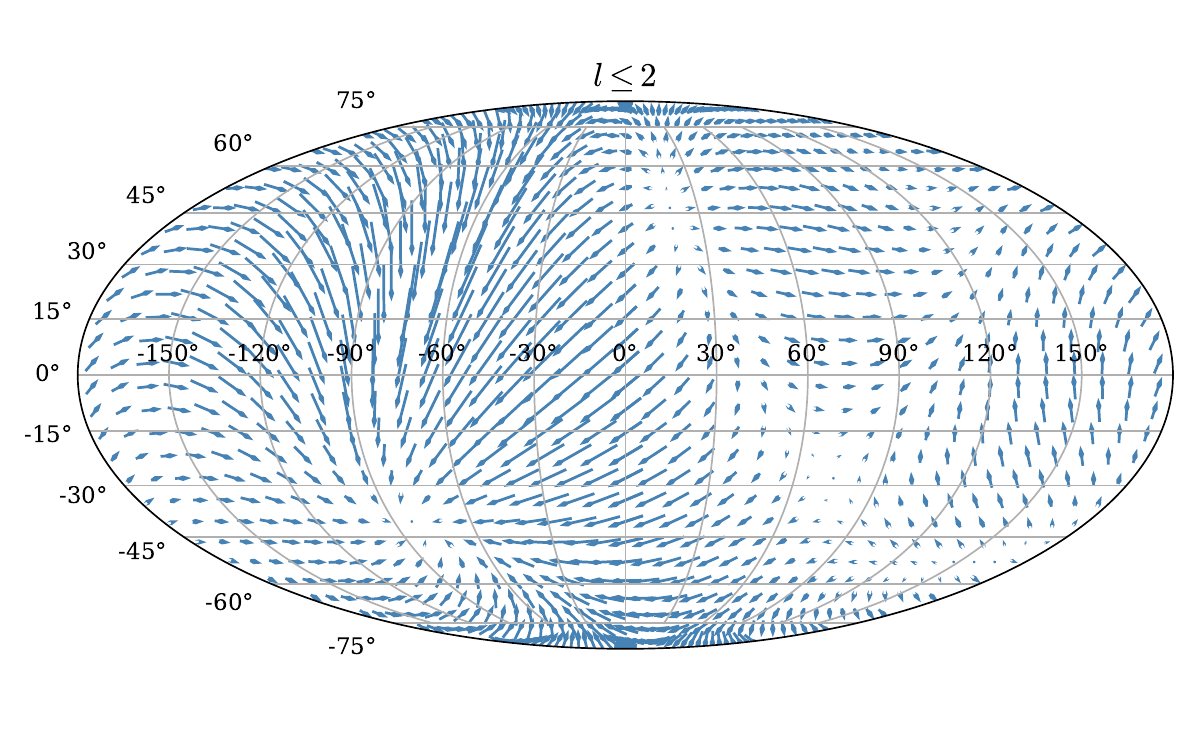}
    \includegraphics[width=0.49\textwidth]{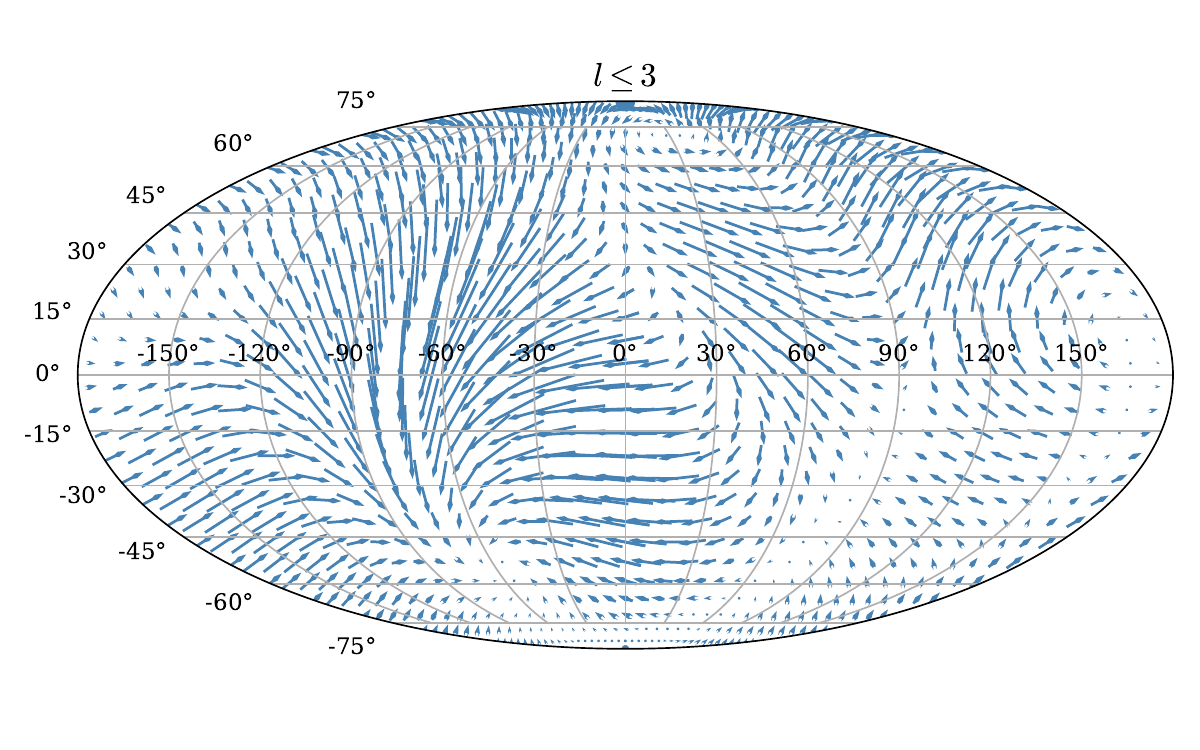}
    \caption{Mollewide maps {of the proper motion field} assoicated with the $l=1$ (top), $l\leq 2$ (middle), and $l\leq 3$ (bottom) VSH coefficients as determined from the full CRF3 dataset of five-parameter sources.  }
    \label{fig:mollewide_maps} 
\end{figure}

{Here we provide a visualization of the low multipoles of the QSO position drift signal for the total CRF3 dataset, as determined in Section~\ref{sec:full}. 
We show the Mollewide projection of the $l = 1$, $l = 2$, and $l = 3$ proper motion field contributions in  Fig.~\ref{fig:mollewide_maps}, where the VSH coefficients used are the Bayesian means determined in Section~\ref{sec:full}.   }
{At each order, we see that higher order moments introduce flow patterns at the associated angular scale.  }

\section{Convergence of the redshift dependence analysis}
\label{Appendix B}

\begin{figure}
    \includegraphics[width=0.48\textwidth]{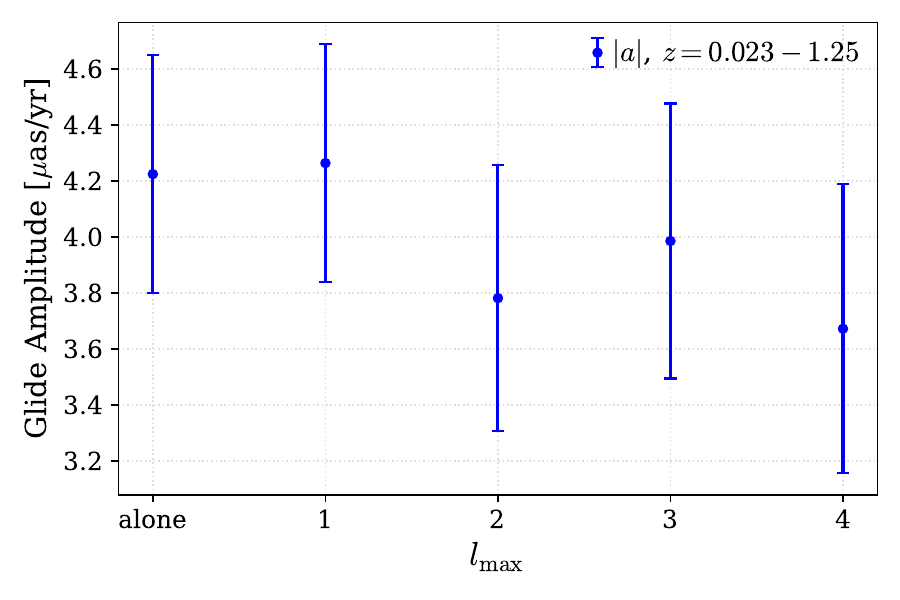}
    \includegraphics[width=0.48\textwidth]{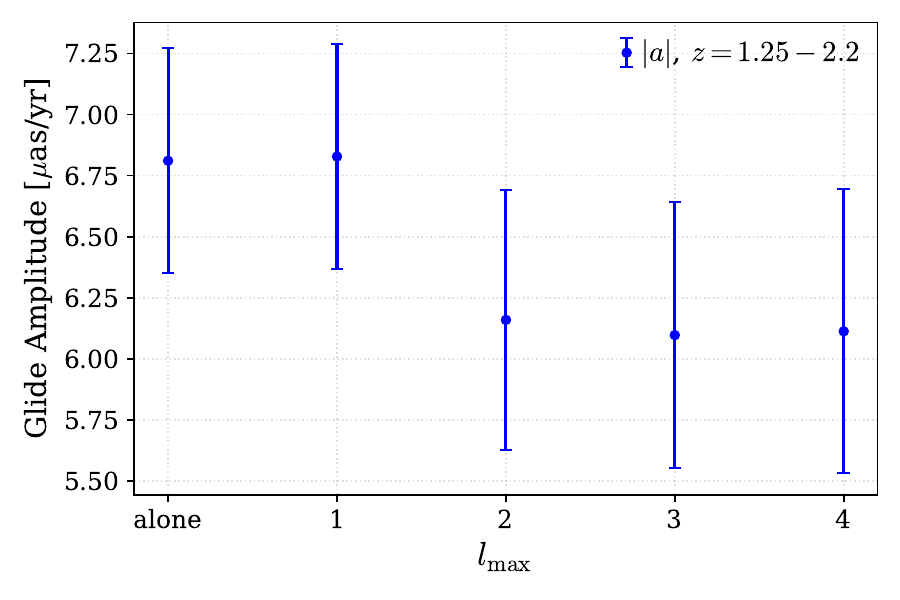}
    \centering\includegraphics[width=0.48\textwidth]{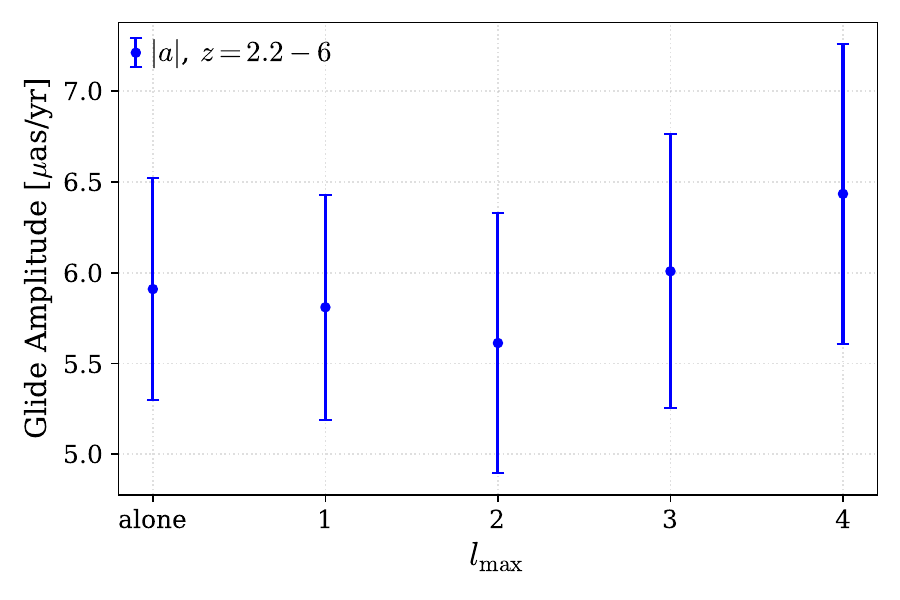}
    \caption{Convergence of the glide amplitude {with truncation of the VSH expansion $l_{\max}$ for each} redshift bin {used in the main analysis}.}
    \label{fig:convergence}
\end{figure}

{Here, we analyze the convergence of the results for the glide vector with $l_{\max}$ in the three redshift bins considered in Section~\ref{sec:zdep}, similar to the convergence analysis for the full CRF3 sample shown in Fig.~\ref{fig:acceleration_rotation_components}. 
We focus on the amplitude of the glide vector, in order to make a direct link to the amplitude estimates in Section~\ref{sec:zdep}, which are our main interest in our redshift dependence analysis.  } 

{The results in Fig.~\ref{fig:convergence} indicate stability of the results to the choice of cutoff, $l_{\max}$, of the VSH expansion for all redshift bins, with dependence on $l_{\max}$ after $l_{\max} \geq 2$ generally being well within the statistical error bars. }   
{We find similar results in our complete analysis of the convergence of all of the dipole and quadrupole components: the results appear robust with  $l_{\max}$, and convergence generally appears to have been reached at a good precision for $l_{\max} = 4$. 
}

These results give confidence that our results for the redshift evolution obtained in Section~\ref{sec:zdep} are reliable and not due to spurious effects caused by leakage of higher-order $l > 4$ multipoles. In particular, we conclude that the statistically-significant result for the redshift evolution of the glide vector amplitude as seen in the top left panel of  Fig.~\eqref{fig:redshift_dependence} is robust.

\end{document}